\documentclass[conference]{IEEEtran}

\usepackage{cite}
\usepackage{amsmath,amssymb,amsfonts}
\usepackage{algorithmic}
\usepackage{graphicx}
\usepackage{textcomp}
\usepackage[dvipsnames]{xcolor}
\usepackage{multirow}
\usepackage[hidelinks, bookmarks=false]{hyperref}
\usepackage{balance}

\usepackage{array,tabularx,booktabs,makecell,pifont}
\renewcommand{\arraystretch}{1}
\newcolumntype{L}[1]{>{\raggedright\arraybackslash}p{#1}}
\newcolumntype{C}[1]{>{\centering\arraybackslash}p{#1}}
\newcommand{\cmark}{\ding{51}}
\newcolumntype{Y}{>{\centering\arraybackslash}X}

\usepackage{siunitx}
\newcommand{\dspu}[2]{%
  \num[group-separator={,},group-minimum-digits=4]{#1}\,(\num{#2}\,\%)%
}

\definecolor{GrayCustom}{gray}{0.4}
\newcommand{\grayit}[1]{%
  {\color{GrayCustom}\tablenum[group-separator={,},group-minimum-digits=4]{#1}}%
}

\makeatletter
\def\ps@IEEEtitlepagestyle{%
  \def\@oddfoot{%
    \hbox to \textwidth{%
      \hfil
      \parbox{\textwidth}{\centering\scriptsize
      \textcopyright~2025 IEEE. Personal use of this material is permitted. Permission from IEEE must be obtained for all other uses, in any current or future media, including reprinting/republishing this material for advertising or promotional purposes, creating new collective works, for resale or redistribution to servers or lists, or reuse of any copyrighted component of this work in other works.}%
      \hfil
    }%
  }%
  \def\@evenfoot{}%
}
\makeatother

\title{Instruction-Based Coordination of Heterogeneous Processing Units for Acceleration of DNN Inference}

\author{\IEEEauthorblockN{Anastasios Petropoulos and Theodore Antonakopoulos}
\IEEEauthorblockA{Dept. of Electrical and Computer Engineering, University of Patras, Patras, Greece\\
\mbox{a.petropoulos@ece.upatras.gr, antonako@upatras.gr}}}

\begin{document}
\maketitle

\begin{abstract}
This paper presents an instruction-based coordination architecture for Field-Programmable Gate Array (FPGA)-based systems with multiple high-performance Processing Units (PUs) for accelerating Deep Neural Network (DNN) inference. This architecture enables programmable multi-PU synchronization through instruction controller units coupled with peer-to-peer instruction synchronization units, utilizing instruction types organized into load, compute, and store functional groups. A compilation framework is presented that transforms DNN models into executable instruction programs, enabling flexible partitioning of DNN models into topologically contiguous subgraphs mapped to available PUs. Multiple deployment strategies are supported, enabling pipeline parallelism among PUs and batch-level parallelism across different PU subsets, with runtime switching among them without FPGA reconfiguration. The proposed approach enables design space exploration, supporting dynamic trade-offs between single-batch and multi-batch performance. Experimental results on ResNet-50 demonstrate notable compute efficiency, up to $98\%$, and throughput efficiency gains, up to $2.7\times$, over prior works across different configurations.
\end{abstract}


\vspace{-0.25cm}
\section{Introduction and Related Work}
\vspace{-0.05cm}
The advancement of Deep Neural Networks (DNNs) has driven the demand for specialized hardware acceleration, with Field-Programmable Gate Arrays (FPGAs) considered as promising accelerators due to their reconfigurability and parallel processing capabilities for DNN inference \cite{nechi2023fpga}. Numerous automated design frameworks have emerged for DNN-to-FPGA acceleration \cite{zeng2021unified, ma2018automatic, zhang2018dnnbuilder, ye2020hybriddnn, d2022xdnn, zhou2024pipefuser, zhang2020dnnexplorer, basalama2023flexcnn, xing2019dnnvm, li2025hardware}, typically following two architectural strategies: unified Processing Unit (PU) architectures and heterogeneous PU architectures, as described in \cite{cai2022deepburning, zhou2024pipefuser}.

Unified PU architectures employ generic accelerators that execute DNN layers sequentially. These designs typically utilize systolic array (SA)-based PUs with DSP48E2 units \cite{amd_dsp48e2_guide}, operating at high clock frequencies, to achieve high throughput \cite{shi2020ftdl, d2022xdnn, Fan2022Accel, petropoulos2025}. Recent unified accelerators target diverse models, such as architectures that support both attention and convolutional (Conv) operations \cite{li2023unified} or versatile designs that handle Convolutional Neural Networks (CNNs), Graph Neural Networks (GNNs), and Vision Transformers (ViTs) \cite{zhang2024visionagile}. Flexible architectures have also emerged for arbitrary-kernel convolutions \cite{Wu2024Amoeba} and the exploitation of dynamic parallelism in each DNN layer \cite{dai2024dcp}. Although multiple PU instantiations enable batch-level parallelism \cite{d2022xdnn, petropoulos2025, amd_pg367}, these approaches process layers within individual PUs, leaving inter-PU pipeline opportunities unexploited. Thus, the sequential nature inherently limits single-batch performance, requiring large batch sizes to achieve high throughput.

Heterogeneous architectures take a different approach, exploiting dedicated PUs for subsets of DNN layers \cite{xiao2017exploring, shen2017maximizing, wu2019compute, chan2019partitioning} or even for each layer \cite{doumet2024h2pipe}, enabling pipelining across entire models. Some designs also focus on low-latency inference through optimized heterogeneous accelerators \cite{wei2018tgpa}. In addition, hybrid approaches have emerged that combine elements of both strategies to balance flexibility and performance. In this context, segment-grained pipeline architectures share PUs across DNN model segments \cite{cai2022deepburning}. In contrast, others fuse layer groups into pipeline stages \cite{zhou2024pipefuser} or adopt heterogeneous PUs for initial layers and uniform PUs for the remaining layers \cite{zhang2020dnnexplorer}. SA-based implementations are also commonly found in heterogeneous architectures \cite{wu2019compute, samajdar2019scaling, chan2019partitioning}.

Instruction Set Architectures (ISAs) provide runtime programmability for different DNN models without FPGA reconfiguration \cite{abdelfattah2018dla, xing2019dnnvm, yu2019opu, ye2020hybriddnn, d2022xdnn, amd_pg367}. However, existing instruction-based approaches focus on single-PU control, and extending these concepts to multi-PU coordination presents challenges in distributed execution and inter-PU synchronization. Multi-PU approaches typically operate either as batch-level parallelism with independent PU processing of complete models, or as pipeline execution among PUs for single batches. These execution schemes are generally fixed at design time, limiting flexibility under dynamic workload conditions.

This work introduces an instruction-based coordination architecture that enables the synchronization of multiple PUs without requiring FPGA reconfiguration. Our approach employs an Instruction Controller Unit (ICU) within each PU and peer-to-peer Instruction Synchronization Units (ISUs) that manage inter-PU synchronization through hardware-based request-acknowledgment mechanisms. The key innovation lies in expressing coordination patterns within instruction sequences that execute in each PU. Also, via instruction updates, the architecture can enable runtime switching between pipeline parallelism for single-batch acceleration and hybrid parallelism, which combines pipeline and batch-level strategies.

In addition, we developed a comprehensive DNN compilation framework that transforms DNN models into instruction programs optimized for multi-PU execution. In particular, it encompasses model processing with hardware-aware fusion, node-to-PU partitioning with weight transfer scheduling, pipeline memory management, and instruction generation. This compilation approach, along with the proposed architecture, enables systematic Design Space Exploration (DSE) across diverse deployment strategies for selecting deployment configurations, allowing runtime adaptation based on application requirements and constraints.

Our contributions include: 1) an instruction-based coordination architecture enabling programmable multi-PU synchronization; 2) a compilation framework transforming DNN models into instruction sequences and supporting flexible parallelism strategies; 3) a DSE methodology for the system architecture for single-batch and multi-batch configurations; and 4) experimental validation on the ResNet-50 across different deployment configurations, demonstrating notable compute and throughput efficiency gains over prior works.

The remainder of this paper is organized as follows. Section II presents the hardware foundation, covering the baseline heterogeneous PU architecture and the custom ISA. Section III details the multi-PU coordination architecture, including the ICU and ISU components. Section IV describes the DNN compilation framework, while Section V presents the implementation details, the DSE methodology, and the performance evaluation for ResNet-50.

\begin{table*}[!ht]
\centering
\caption{ISA Overview}
\label{tab:ICU}
\vspace{-0.25cm}
\scriptsize
\makebox[\textwidth]{%
    \begin{minipage}[t]{0.47\textwidth}
        \centering
        \textbf{(a) Instruction Types} \\[0.4ex]
            \begin{tabularx}{\columnwidth}{@{}
                  L{1.6cm}  
                  C{0.4cm}  
                  C{0.4cm}  
                  C{0.4cm}  
                  X          
                @{}}
                \toprule
                \textsc{Type}\textsuperscript{1}
                  & \multicolumn{3}{c}{\textsc{ICU Groups}}
                  & \textsc{Description; Key Fields}\textsuperscript{2}\\
                \cmidrule(lr){2-4}
                  & \textsc{LD} & \textsc{CP} & \textsc{ST} &\\
                \midrule

                \makecell[lt]{\textbf{ProgCtrl}\\{\tiny\itshape PRG\_PRM}}
                  & \cmark & \cmark & \cmark
                  & Control program loops; \texttt{NR, ICU\_BA}\\
                \specialrule{0.03pt}{0.3pt}{0.3pt}
                  
                \makecell[lt]{\textbf{Config}\\{\tiny\itshape *\_PRM}}
                  & \cmark & \cmark & \cmark
                  & Prepare PU before ADM; \texttt{stride\_pattern, IM2COL, URAM\_addr}\\
                \specialrule{0.03pt}{0.3pt}{0.3pt}
            
                \makecell[lt]{\textbf{DataMove}\\{\tiny\itshape *\_ADM}}
                  & \cmark & \cmark & \cmark
                  & ADM transfers control; \texttt{CUR\_BA, LEN}. \texttt{CUR\_BA} latched for successor \textbf{AddrCyc}.\\
                \specialrule{0.03pt}{0.3pt}{0.3pt}
            
                \makecell[lt]{\textbf{AddrCyc}\\{\tiny\itshape CYCLE\_ADDR}}
                  & \cmark & \cmark & \cmark
                  & Cyclic addressing; \texttt{BA, AOFFS, NC, IC}\\
                \specialrule{0.03pt}{0.3pt}{0.3pt}
            
                \makecell[lt]{\textbf{Sync}\\{\tiny\itshape SEND/WAIT\_REQ/ACK}}
                  & \cmark &        & \cmark
                  & Inter-ICU coordination; \texttt{DST/SRC\_PID, BID, BASE\_BID, NC, IC}\\
                \specialrule{0.03pt}{0.3pt}{0.3pt}
            
                \makecell[lt]{\textbf{Compute}\\{\tiny\itshape GEMM}}
                  &        & \cmark &  
                  & PU operations; \texttt{ReLu, Rounds, Scales, Add\_enable}\\
                  
                \bottomrule
          \end{tabularx}
    \end{minipage}%
    \hspace{0.02\textwidth}%
    \begin{minipage}[t]{0.51\textwidth}
        \centering
        \textbf{(b) Key Instruction Type Algorithms \& Write-Back} \\[0.4ex]
            \begin{tabularx}{\columnwidth}{@{}
              L{0.9cm}  
              X         
            @{}}
            \toprule
            \textsc{Type} & \textsc{Pseudocode \& Write-Back}\textsuperscript{3}\\
            \midrule

           \textbf{ProgCtrl}
              & \texttt{if (NR == 0): $\infty$-loop: run\_until(PRG\_END); jump(ICU\_BA)}\\
              & \texttt{else: repeat(NR): run\_until(PRG\_END); jump(ICU\_BA)}\\
             \specialrule{0.03pt}{0.3pt}{0.3pt}
              
            \textbf{AddrCyc}
              & \texttt{if (IC == 0): IC, CUR\_BA = NC, BA}\\
              & \texttt{else: IC, CUR\_BA = IC-1, CUR\_BA+AOFFS}\\
              & \textit{Updates:} predecessor \texttt{CUR\_BA}, current \texttt{IC}\\
            \specialrule{0.03pt}{0.3pt}{0.3pt}

            \textbf{Sync}
              & \texttt{if (NC == 0): BID = BID} \textcolor{gray}{\scriptsize // bypass}\\
              & \texttt{elif (IC == 0): BID, IC = BASE\_BID, NC}\\
              & \texttt{else: BID, IC = BID+1, IC-1} \\
              & \textit{Updates:} current \texttt{BID}, \texttt{IC}\\
  
            \bottomrule
            \end{tabularx}
            
        \begin{flushleft}
        \scriptsize
        \textsuperscript{1} ProgCtrl, Config, Compute: static instructions; Others: dynamic ICU BRAM write-back.\\
        \textsuperscript{2} All instructions include \texttt{OPCD}, \texttt{PRG\_END} fields.\\
        \textsuperscript{3} \textit{Updates:} Write-back to ICU BRAM. \texttt{IC} init: \texttt{NC} when loaded offline.\\
        \textsuperscript{4} $\rightarrow$: mandatory sequence. $^*$: optional successor to any \texttt{*\_ADM}.
        \end{flushleft}
    \end{minipage}%
}

\vspace{0.4em}

\noindent\textbf{(c) Basic Instructions in each ICU Group} \\[0.3ex]
\begin{tabularx}{\textwidth}{@{}p{0.5cm}X@{}}
\toprule
\textsc{ICU} & \textsc{Basic Instructions}\textsuperscript{4} \\
\midrule
\textbf{LD} & \texttt{LINEAR\_ADM, (IM2COL\_PRM$\rightarrow$IM2COL\_ADM), (STRIDE\_PRM$\rightarrow$STRIDE\_ADM), SEND\_ACK, WAIT\_REQ, CYCLE\_ADDR$^*$, PRG\_PRM} \\
\addlinespace[0.05em]
\textbf{CP} & \texttt{(URAM\_PRM$\rightarrow$WEIGHTS\_ADM), (RES\_ADD\_STRIDE\_PRM$\rightarrow$RES\_ADD\_STRIDE\_ADM), RES\_ADD\_ADM, CYCLE\_ADDR$^*$, GEMM, PRG\_PRM} \\
\addlinespace[0.05em]
\textbf{ST} & \texttt{LINEAR\_ADM, (STRIDE\_PRM$\rightarrow$STRIDE\_ADM), SEND\_REQ, WAIT\_ACK, CYCLE\_ADDR$^*$, PRG\_PRM} \\
\bottomrule
\end{tabularx}
\vspace{-0.45cm}
\end{table*}

\vspace{-0.1cm}
\section{Hardware Foundation}
\vspace{-0.05cm}
In this section, the hardware foundation for multi-PU coordination is presented, covering the baseline heterogeneous design and the custom ISA that enables programmable coordination and execution among PUs.

\vspace{-0.1cm}
\subsection{Baseline Architecture} \label{sec:base_arch}
\vspace{-0.05cm}
This work builds upon a heterogeneous multi-PU architecture designed for FPGA-based DNN acceleration \cite{petropoulos2025}, providing the hardware baseline enhanced in this work with instruction-based coordination mechanisms for programmable inter-PU synchronization.

\begin{figure}[!t]
\centering
\includegraphics[width=1.0\columnwidth]{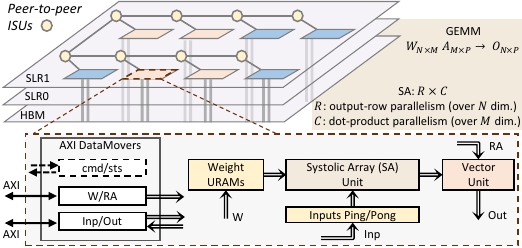}
\vspace{-0.7cm}
\caption{System architecture of multiple PUs. Inset: PU architecture \cite{petropoulos2025}.}
\label{fig:arch_overview}
\vspace{-0.5cm}
\end{figure}

In the baseline architecture, five $\text{PU}_{1\mathrm{x}}$ and five $\text{PU}_{2\mathrm{x}}$ units are instantiated across the Super Logic Regions (SLRs) of the FPGA. $\text{PU}_{1\mathrm{x}}$ and $\text{PU}_{2\mathrm{x}}$ units are configured as 64$\times$4 and 64$\times$8 SAs, respectively, with the latter exhibiting twice the computational performance. Both PU types support Conv and Fully Connected (FC) layers via General Matrix-Matrix Multiplication (GEMM) operations. Fig.~\ref{fig:arch_overview} illustrates the multi-PU system organization, showing distributed PUs across different FPGA regions abstractly, with ISUs forming the coordination infrastructure introduced in this work for orchestrating the dataflow across PUs during DNN inference.

The PU, as shown in Fig.~\ref{fig:arch_overview} inset, comprises three functional blocks. The pre-processing block contains two AXI DataMover (ADM) modules that interface with the High-Bandwidth Memory (HBM) and utilize Block RAM (BRAM)-based ping-pong activation buffers for continuous data streaming. The main block features the SA for GEMM operations, utilizing UltraRAMs (URAMs) to store weights and biases at twice the system clock (sys\_clk) frequency, denoted as dsp\_clk. The post-processing block comprises activation functions, vector units that support residual additions, and a wave reorder buffer that manages out-of-order systolic waves, enabling about $98$\% performance efficiency on ResNet-50.

\vspace{-0.1cm}
\subsection{Instruction Set Architecture} \label{sec:ISA}
\vspace{-0.05cm}
The ICU architecture employs a custom ISA, designed to support PU operations and to coordinate dataflow execution across multiple PUs. Instructions are organized into six distinct types that collectively enable configurable control of memory transfers, computations, and inter-PU synchronization. This instruction-based approach enables runtime reconfiguration without hardware changes, supporting different DNN models and PUs configurations by updating only the PU instructions.

The instruction set covers three ICU groups (\textit{Load, \mbox{Compute}, Store}), each optimized for specific dataflow functions, decoupling of memory accesses from computational tasks, and enabling prefetching and overlapping of data transfers with SA operations. The \mbox{\textit{Load (LD)}} group handles input activation data management, including linear, Image to Column (IM2COL) transform, and stride-patterned data movement capabilities, along with acknowledgment transmission and request awaiting for synchronization. The \mbox{\textit{Compute (CP)}} group focuses on GEMM operations, weights management, and residual shortcut activation data transfers, supporting both linear and stride-patterned transfers. The \mbox{\textit{Store (ST)}} group manages output activation data transfers and complements the \textit{LD} group's synchronization capabilities by providing transmission request and acknowledgment awaiting functions.

Each ICU group has its dedicated dual-port BRAM for storing instruction programs, with program rounds representing complete execution cycles that iterate through instructions in sequential order until reaching the end of the program, i.e., an instruction with the \texttt{PRG\_END} field set. All instructions use 64-bit instruction length and include operation code (\texttt{OPCD}) and \texttt{PRG\_END} fields for consistent decoding. In addition, they are classified into two categories, based on their runtime behavior: \textit{static instructions} remain unchanged during program execution, while \textit{dynamic instructions} update their state through write-back mechanisms that modify the respective ICU BRAM parameters.

In Table~\ref{tab:ICU}(a), the core instruction types are detailed and described below, with each type serving a specific role in PU operations. Table~\ref{tab:ICU}(b) presents the dynamic state update algorithms, while Table~\ref{tab:ICU}(c) shows the basic instructions within each ICU group.
\begin{itemize}[
    \setlength{\IEEElabelindent}{\dimexpr-\labelwidth-\labelsep}
    \setlength{\itemindent}{\dimexpr\labelwidth+\labelsep}
    \setlength{\listparindent}{\parindent}
]
    \item \textit{ProgCtrl}: It manages program flow and looping behavior for all ICU groups. The rounds parameter (\texttt{NR}) controls the instruction program behavior: zero enables infinite loops until reset, otherwise, it executes the specified rounds with the instruction pointer jumping to the designated base address (\texttt{ICU\_BA}) at the end of each round. This static instruction type is essential for multiple inference rounds in DNN applications.
    \item \textit{Config}: It prepares PUs by setting parameters for stride memory access patterns, IM2COL operations, and URAM addressing. These static instructions establish the operational context for subsequent \textit{DataMove} instructions.
    \item \textit{DataMove}: It controls ADM operations for memory transfers between HBM and the PU on-chip buffers. It specifies the current base address (\texttt{CUR\_BA}) and transfer length (\texttt{LEN}), with \texttt{CUR\_BA} latched for successor \textit{\mbox{AddrCyc}} instructions to enable dynamic address management.
    \item \textit{AddrCyc}: It implements cyclic addressing for efficient memory utilization during inference. It operates on base address (\texttt{BA}), offset (\texttt{AOFFS}), cycles (\texttt{NC}), and iteration counter (\texttt{IC}), with the algorithm [see Table~\ref{tab:ICU}(b)] managing address progression and counter updates across program rounds.
    \item \textit{Sync}: It coordinates buffer synchronization between cooperating PUs via request-acknowledgment messages. These instructions are exclusive to \textit{LD} and \textit{ST} groups, and include PU identifiers (\texttt{PID}), buffer identifiers (\texttt{BID}), and cyclic parameters. The algorithm [see Table~\ref{tab:ICU}(b)] provides bypass, reset, and cyclic increment operations.
    \item \textit{Compute}: It controls SA and vector operations in the PUs for the \textit{CP} group, and configures activation functions, quantization scaling factors, and residual addition.
\end{itemize}

This instruction-level approach eliminates external control logic, enabling coordination sequences to be expressed as instruction programs that execute within each ICU. As the PU designs advance, the ISA can accommodate new instruction types, allowing the expansion of computational capabilities without requiring fundamental architectural changes.

\vspace{-0.1cm}
\section{Multi-PU Coordination Architecture} \label{sec:ICU_ISU}
\vspace{-0.05cm}

\begin{figure*}[!t]
\centering
\includegraphics[width=1.0\textwidth]{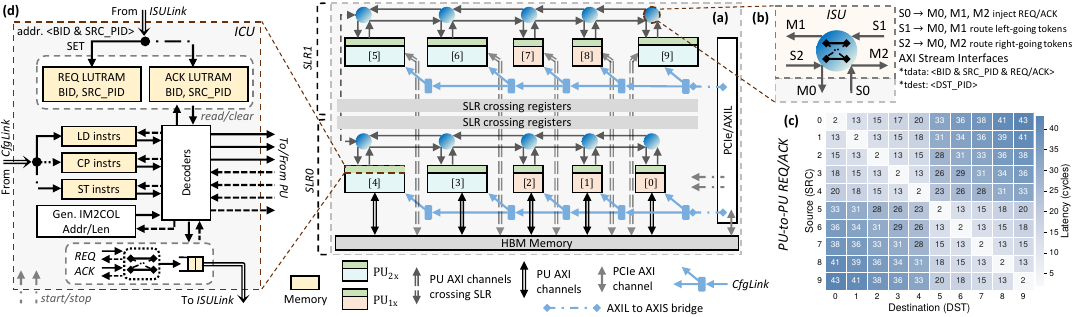}
\vspace{-0.7cm}
\caption{Instruction-based coordination architecture: (a) System architecture with PUs distribution. (b) ISU. (c) Control token PU-to-PU latencies. (d) ICU.}
\label{fig:arch_icu_isu}
\vspace{-0.5cm}
\end{figure*}

Building upon the ISA described in Section~\ref{sec:ISA}, here we detail how the coordination architecture enables programmable multi-PU execution. The approach uses an ICU in each PU for local instruction execution and coordination logic, and peer-to-peer ISUs that manage inter-PU flow control via hardware-based control tokens (REQ/ACK messages). Each ICU independently manages the progression of its instruction pointers and states, with the distributed token-based coordination eliminating the need for centralized control.

Fig.~\ref{fig:arch_icu_isu} illustrates the system architecture, showing the PUs organization using the hardware baseline as described in Section~\ref{sec:base_arch}, detailed ICU and ISU architectures, and the measured control token latencies. During system initialization, instruction programs are loaded offline via a cascaded configuration link (CfgLink) that employs AXI4-Lite (AXIL) control registers bridged to a daisy-chained AXI4-Stream (AXIS) interface. This configuration link includes embedded switches that route instruction data, first by \texttt{PID} and then by ICU group, enabling the host to write instruction programs to designated ICU groups in each PU. Once loaded, instruction programs execute upon receiving a start signal from the host.

\vspace{-0.1cm}
\subsection{Instruction Synchronization Network}
\vspace{-0.05cm}
The ISUs network implements a distributed switch fabric that routes single-beat control tokens between PUs using AXIS channels, as shown in Fig.~\ref{fig:arch_icu_isu}(b). Each ISU contains an AXIS switch with register slices and dedicated master-slave interfaces. These comprise slave interfaces for local ISU injection (S0) and bidirectional control token forwarding (S1, S2), along with master interfaces for local ISU delivery (M0) and directional routing (M1, M2). The switch employs fixed routing tables and utilizes one-transfer round-robin arbitration to resolve contention when multiple tokens compete for the same port.

Control tokens carry synchronization information in a compact AXIS format with TDATA fields encoding the \texttt{BID}, source PU ID (\texttt{SRC\_PID}), and REQ/ACK type. In contrast, TDEST fields specify the destination PU (\texttt{DST\_PID}) for routing decisions. The static routing tables enable deterministic latency characteristics captured in the measured latency matrix, as shown in Fig.~\ref{fig:arch_icu_isu}(c). This matrix is derived under nominal conditions with a single token in transit at the sys\_clk rate, measuring the cycle count from token transmission to destination delivery. The results reveal the architectural impact of the SLR boundary crossing, which adds a $13$-cycle penalty to cross-SLR paths. In contrast, same-SLR hops incur $2$\,\text{--}\,$3$ cycles, and same-PU delivery bypasses the switch fabric, resulting in a $2$-cycle latency. Consequently, given that PU execution operates in the hundreds of microseconds timescale for most DNN models, while control tokens complete in sub-microseconds, potential contention effects are negligible for the intended coordination requirements.

\vspace{-0.1cm}
\subsection{Instruction Controller Unit}
\vspace{-0.05cm}
Each PU contains a dedicated ICU that executes instruction programs from three independent dual-port BRAMs corresponding to the ICU groups, as depicted in Fig.~\ref{fig:arch_icu_isu}(d). The ICU incorporates separate instruction decoders for each group, enabling parallel execution of \textit{LD}, \textit{CP}, \textit{ST} operations within a PU. This execution scheme enables the decoupling of memory accesses from computational operations, allowing overlapped pipelining and simplifying instruction generation. 

Critical to the coordination mechanism are the REQ and ACK LUTRAMs that store the synchronization states for inter-PU flow control. These LUTRAMs are addressed using the \texttt{BID} and \texttt{SRC\_PID} fields, enabling the ICU to track multiple outstanding synchronization messages. The \texttt{BID} parameter enables pipeline coordination by distinguishing between multiple in-flight data transfers in balanced and unbalanced pipeline configurations. Its cyclic behavior ensures buffer management without race conditions in dependency chains that span multiple PUs, as demonstrated in Section~\ref{sec:multi_PU_ex}.

The REQ/ACK messages are issued via complementary instruction pairs specified in the \textit{Sync} instruction type [see Table~\ref{tab:ICU}(a)]. The \texttt{SEND\_REQ/ACK} instructions transmit synchronization messages to destination PUs, whereas the \texttt{WAIT\_REQ/ACK} instructions monitor the related LUTRAMs for control tokens from source PUs. When a token is received through the ISULink interface (M0 connection from local ISU), the ICU extracts the REQ/ACK type from the token data and updates the corresponding LUTRAM address. Subsequently, \texttt{WAIT} instructions act as barriers by polling the LUTRAMs until synchronization is satisfied, after which the associated entries are cleared and execution proceeds.

Outgoing synchronization messages are issued by \texttt{SEND} instructions and forwarded through a multiplexer to the ISULink interface (S0 connection to local ISU). A small FIFO buffer at this interface enables the ICU decoder FSMs to proceed after issuing instructions, without blocking until the ISU network becomes available. This maintains instruction execution flow and avoids stalls in the local instruction pipeline.

\vspace{-0.1cm}
\subsection{Two-PU Pipeline Example} \label{sec:multi_PU_ex}
\begin{figure}[!t]
\centering
\includegraphics[width=1.0\columnwidth]{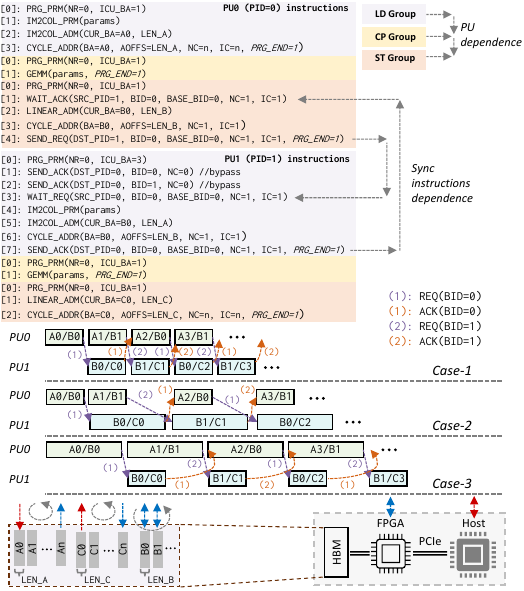}
\vspace{-0.7cm}
\caption{Two-PU pipeline coordination example: instructions/timing overview.}
\label{fig:pu_example}
\vspace{-0.5cm}
\end{figure}

\vspace{-0.05cm}
To demonstrate these coordination capabilities, we present a two-PU pipeline example, mapping two Conv layers (transformed to GEMM) to PU0 and PU1, respectively, creating a producer-consumer dependency chain, where PU0 generates intermediate activations consumed by PU1. Fig.~\ref{fig:pu_example} illustrates the instruction execution, buffer management, and synchronization across balanced and unbalanced pipeline scenarios. The memory hierarchy employs three tensor buffer categories: \textit{A}-regions store input activations, \textit{B}-regions provide intermediate activations storage with ping-pong buffering (\texttt{BID=0}, \texttt{BID=1}), and \textit{C}-regions store outputs. Both \textit{A} and \textit{C}-regions use cyclic access patterns across \textit{n} HBM regions in each, whereas \textit{B}-regions enable overlapped execution while maintaining data dependencies.

In Fig.~\ref{fig:pu_example} (top part), the instruction programs are shown, where PU0's \textit{ST} group uses \texttt{WAIT\_ACK/SEND\_REQ} instructions, while PU1's \textit{LD} group issues the reciprocal \texttt{WAIT\_REQ/SEND\_ACK} handshakes. Pipeline initialization relies on ACK bypass in PU1's \textit{LD} instructions at addresses \{1, 2\}, pre-authorizing PU0 to use both \textit{B0}, \textit{B1} buffers by setting the appropriate ACK LUTRAM addresses before execution. Also, \texttt{CYCLE\_ADDR} instructions manage the cyclic patterns: \textit{n} cycles for \textit{A/C}-regions versus ping-pong alternation for \textit{B0/B1} intermediate buffers (\texttt{NC=1} creates a two-region cycle), reflecting different buffer management requirements.

In Fig.~\ref{fig:pu_example} (bottom part), balanced pipeline operation emerges when both PUs achieve comparable throughput (Case-1). Following the ACK bypass warm-up, steady-state execution proceeds as PU1 processes round \textit{N-1} using buffer \texttt{BID=X} while PU0 executes round \textit{N}, writing to buffer \texttt{BID=1-X}, thereby eliminating memory conflicts. As for the intra-PU pipelines, they operate with \textit{LD$\rightarrow$CP$\rightarrow$ST} dependencies during warm-up, transitioning to parallel execution as PU activation buffers allow overlapped memory and computation operations.

On the contrary, when PU1 operates at half the throughput of PU0 (Case-2), bottlenecks manifest as extended ACK wait intervals for PU0. After initial bypass rounds, PU0's \textit{ST} group blocks await ACK messages from PU1, filling on-chip output buffers and triggering \textit{ST$\rightarrow$CP$\rightarrow$LD} back-pressure propagation that throttles PU0 to match PU1's rate. Case-3 reverses this dynamic, with PU1 awaiting extended periods for REQ messages, which indicate data availability. In this case, ACK messages are unnecessary, as PU0's reduced throughput prevents memory contention. Notably, instruction uniformity is maintained across the cases, regardless of performance characteristics.

Cyclic buffering in the \textit{A} and \textit{C}-regions enables concurrent PCIe transfers with PU execution. The host writes new input batches to available \textit{A}-regions while reading the results from \textit{C}-regions as the PUs execute. Intermediate \textit{B} buffers require only two regions, since the producer-consumer access patterns between PU0 and PU1 guarantee buffer consistency. That scales to deeper pipelines requiring additional \texttt{BID} values proportional to the pipeline depth. Complex topologies with multiple consumers (e.g., PU0 feeding both PU1 and PU2) increase the \texttt{BID} cyclic depth proportionally, when all PUs achieve comparable throughput, while expanding REQ/ACK instructions for synchronization. The required buffer depth for each tensor depends on the pipeline configuration, as detailed in Section~\ref{sec:framework_mem_opt}.

\begin{figure}[!t]
\centering
\includegraphics[width=1.0\columnwidth]{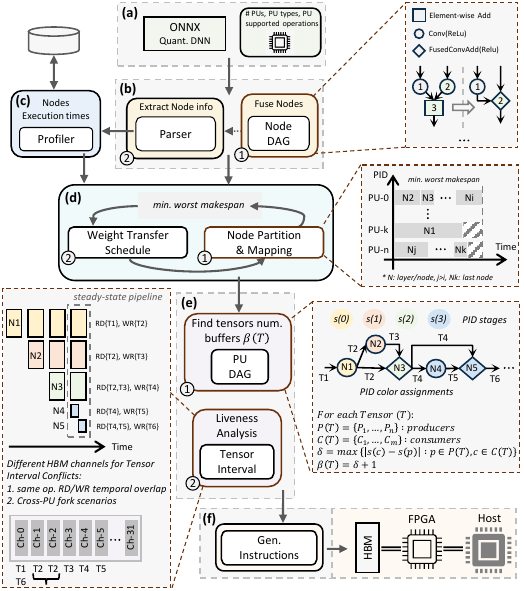}
\vspace{-0.7cm}
\caption{DNN Compilation Framework: (a) DNN model preparation. (b) Node fusion and parser. (c) Node profiler. (d) DNN graph partitioning to PUs and weight transfer scheduling. (e) Tensors buffer optimization and liveness analysis. (f) Instruction generation.}
\label{fig:framework_overview}
\vspace{-0.5cm}
\end{figure}

\vspace{-0.1cm}
\section{DNN Compilation Framework} \label{sec:Framework}
\vspace{-0.05cm}
The instruction-based coordination architecture necessitates a compilation framework that transforms DNN models into executable instruction programs. The proposed multi-phase process, as illustrated in Fig.~\ref{fig:framework_overview}, from model processing through optimization to instruction generation, is described below.

\vspace{-0.1cm}
\subsection{DNN Model Processing and Profiling}
\vspace{-0.05cm}
The proposed framework starts with quantized DNN models in ONNX format \cite{onnx}, utilizing 8-bit quantization with power-of-two scaling factors for compatibility with the PU architecture, as shown in Fig.~\ref{fig:framework_overview}(a). Also, it incorporates system details, i.e., available PU types and supported operations. Using the tile-based execution model \cite{petropoulos2025}, the nodes are partitioned into computational tiles matching the first SA dimension of each mapped PU, enabling node-to-PU assignments and tile-level weight management, as described in Section~\ref{sec:partition_schedule}.

Node fusion adapts DNN graphs to exploit the PU hardware capabilities, while preserving computational correctness, generating the node Directed Acyclic Graph (DAG) [Fig.~\ref{fig:framework_overview}(b1)]. As shown in its inset, Conv layers with subsequent element-wise addition fuse into a \textit{FusedConvAdd(ReLU)} node, as PUs support residual shortcut additions in their dataflow, while the other Conv layer remains unchanged. In addition, the activation functions are integrated with preceding operations. These fusion opportunities can be extended as the PU architecture evolves to support additional hardware-aware optimization patterns and new operations.

The parser extracts node information from the transformed DAG, including weights, biases, quantization scales, dimensions, and operation types [Fig.~\ref{fig:framework_overview}(b2)]. Specifically, it organizes dependency information into structured representations, capturing producer-consumer relationships, and assigns tensor identifiers to all data paths for the next framework phases.

To guide partitioning decisions, execution time profiling is applied under conflict-free conditions, with weights pre-loaded in URAMs, eliminating the memory transfer overhead [Fig.~\ref{fig:framework_overview}(c)]. The measured execution times represent complete node processing, from activation fetching from HBM through computation to output storage. These provide the required performance metrics while abstracting weight loading complexities, which are handled separately in the next phase.

\vspace{-0.1cm}
\subsection{Partitioning and Scheduling} \label{sec:partition_schedule}
\vspace{-0.05cm}
This phase addresses two challenges: assigning nodes to PUs and scheduling weight transfers. The first utilizes dynamic programming to partition the Node DAG topological order into contiguous subgraphs mapped to PUs, considering $\text{PU}_{1\mathrm{x}}$ and $\text{PU}_{2\mathrm{x}}$ heterogeneity and employing the profiled execution times to minimize the maximum completion time across PUs, as shown in Fig.~\ref{fig:framework_overview}(d1).

Weight transfer scheduling addresses the constraint that node tiles within assigned subgraphs often require more weight data than the available URAMs capacity on each PU [Fig.~\ref{fig:framework_overview}(d2)]. Inspired by SMOF \cite{toupas2024}, which fragments weights at design-time and streams chunks at runtime under bandwidth constraints, our approach analyzes tile loading and execution times for timing-based scheduling. Correspondingly, the tile weights are split into chunks, with part of them allocated offline in URAMs, while the rest of them are loaded dynamically from HBM during execution. Dynamic chunks from subsequent tiles are scheduled for loading during the current tile's execution to conceal transfer latency within the execution time.

A greedy deficit-based strategy prioritizes tiles with the highest deficit (chunks causing execution stalls after accounting for overlap hiding opportunities) for offline allocation. The method iteratively allocates chunks to the most deficit-prone tiles, using a priority-based selection, until the capacity constraint is satisfied. In particular, the memory capacity constraint ensures that both statically allocated chunks and the worst-case concurrent dynamic chunks (i.e., the maximum simultaneous memory requirements of two adjacent tiles) do not exceed the total URAMs capacity. Regarding the dynamic chunks, these are evicted after their tile execution completes, releasing storage capacity for subsequent tiles. Although the optimization challenges in this subsection could be considered jointly, the current implementation treats them separately.

\vspace{-0.1cm}
\subsection{Pipeline Memory Optimization} \label{sec:framework_mem_opt}
\vspace{-0.05cm}
For effective pipeline execution, memory management is required to prevent data hazards and to allocate the HBM channels for each DNN tensor. Buffer requirement analysis, as shown in Fig.~\ref{fig:framework_overview}(e1), determines the minimum buffers needed for each tensor to support steady-state pipeline execution without read-after-write and write-after-read dependencies, where consumer and producer PUs must coordinate their accesses. In Fig.~\ref{fig:framework_overview}(e1) inset, the stage-distance method computes the buffer requirements ($\beta(T)$) for each tensor based on the pipeline topology, where stages represent the execution levels in the PU dependency graph. For each tensor, the framework has identified producer-consumer PUs and mapped them to pipeline stages to calculate the maximum stage distance between all producer-consumer pairs. The buffer requirement equals this distance, plus an additional buffer, which enables pipeline operation by allowing producers to write new data while consumers read previously loaded data, thereby preventing pipeline stalls. Longer dependency chains across multiple stages require correspondingly more buffers to prevent access conflicts and maintain synchronization.

Complementing the buffer analysis, tensor liveness analysis identifies HBM channel-sharing opportunities among tensors with non-overlapping access patterns, as detailed in Fig.~\ref{fig:framework_overview}(e2). For that purpose, the pipeline steady-state execution is simulated using the node-to-PU mappings and execution times to track the temporal intervals when tensors are accessed. Thus, tensors with conflicting access patterns, particularly concurrent memory operations of the same type (read-read or write-write), require separate HBM channels to prevent bandwidth bottlenecks \cite{huang2021shuhai}. In addition, cross-PU fork scenarios, where multiple tensors from different PUs are fed to a single consumer (e.g., \textit{FusedConvAddRelu} consuming two tensors), are separately handled. Hence, these tensors are assigned to different HBM channels, ensuring channel sharing decisions do not limit the PUs performance through memory contention.

\vspace{-0.1cm}
\subsection{Instruction Generation}
\vspace{-0.05cm}
The final phase utilizes the optimization results to create executable instruction programs for all the ICUs. Using the instruction types (see Section~\ref{sec:ISA}), the generator produces instruction sequences that embed the optimized node-to-PU assignments and memory management decisions. Specifically, the cyclic buffering patterns are encoded into \texttt{BID} parameters within \textit{Sync} instructions, while cyclic addressing manages transitions between memory buffers for each tensor. The DNN input/output tensors require special handling with cyclic access patterns, coordinated with the PCIe host, following the same principles as the \textit{A} and \textit{C}-regions, as described in Section~\ref{sec:multi_PU_ex}. Conversely, intermediate tensors employ cyclic buffering with \texttt{BID} values rotation for inter-PU coordination. The resulting programs capture both the intra-PU execution logic and the inter-PU coordination, enabling pipeline parallelism across PUs.

\vspace{-0.1cm}
\section{Implementation and Evaluation}
\vspace{-0.05cm}
The proposed architecture was implemented in RTL on an AMD Alveo U50 card, following the heterogeneous PU baseline architecture presented in Section~\ref{sec:base_arch} (5 $\text{PU}_{1\mathrm{x}}$ and 5 $\text{PU}_{2\mathrm{x}}$). Table~\ref{tab:resource_util} details the hardware resource utilization, and as shown, the coordination mechanism introduces minimal resource overhead compared to the baseline implementation.

\begin{table}[!t]
\centering
\caption{System Architecture resource utilization on Alveo U50.}
\label{tab:resource_util}
\vspace{-0.25cm}
\renewcommand{\arraystretch}{0.95}
\scriptsize
\begin{tabularx}{\columnwidth}{@{}l *{4}{Y}@{}}
\toprule
Modules & LUTs & BRAMs & URAMs & DSPs \\
\midrule
PUs                    & 160\,K\,(18.4\,\%) & 200\,(14.9\,\%) & 640\,(100.0\,\%) & 3860\,(64.8\,\%) \\
ICUs, ISUs             & 18.2\,K~(2.1\,\%) & 30~(2.2\,\%) & \text{--} & \text{--} \\
\bottomrule
\end{tabularx}
\vspace{-0.2cm}
\begin{flushleft}
\scriptsize
Clocks: 300 MHz (sys\_clk), 600 MHz (dsp\_clk), 150 MHz (AXIL), 450 MHz (HBM)
\end{flushleft}
\vspace{-0.65cm}
\end{table}

\vspace{-0.1cm}
\subsection{Performance Analysis} \label{sec:perf_analysis}
\vspace{-0.05cm}
The proposed architecture enables flexible deployment strategies from single-batch pipeline execution to multi-batch parallel processing. Here, we evaluate our framework through a DSE methodology and demonstrate how the coordination of multiple PUs unlocks the hybrid parallelism opportunities that combine pipeline and batch-level execution strategies across diverse execution scenarios.

\vspace{-0.15cm}
\begin{figure}[!h]
\centering
\includegraphics[width=1.0\columnwidth]{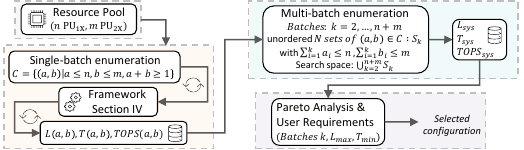}
\vspace{-0.7cm}
\caption{Three-step DSE methodology.}
\label{fig:DSE}
\vspace{-0.15cm}
\end{figure}

The DSE methodology, illustrated in Fig.~\ref{fig:DSE}, provides a systematic approach to performance analysis of DNN models on our architecture, based on the application requirements. In particular, it is a three-step process, which is initiated by enumerating all feasible single-batch configurations, where each configuration (a,b) specifies the assignment of $\text{PU}_{1\mathrm{x}}$ and $\text{PU}_{2\mathrm{x}}$ units respectively, for DNN pipeline execution. In our setup, this enumeration yields 35 distinct single-batch configurations, each generated via the DNN compilation framework (Section~\ref{sec:Framework}), and their resulting performance characteristics are cached. The second step constructs multi-batch schedules by composing all the unordered combinations of single-batch configurations within the constraint of available PU resources. This composition enables hybrid parallelism, where each batch is processed by a subset of PUs exhibiting internal pipeline parallelism. In essence, different batches are processed by separate PU subsets that exhibit batch-level parallelism across them. Each multi-batch schedule is characterized by its aggregated throughput, system latency (determined by the slowest configuration), and the cumulative DSP48E2 Tera Operations per Second (\textit{TOPS}) across all assigned PUs. The final step applies Pareto analysis to identify configurations optimized for different metrics while accommodating application-specific constraints, such as maximum acceptable latency, minimum required throughput, and target batch processing requirements.

\begin{figure}[!t]
\centering
\includegraphics[width=1.0\columnwidth]{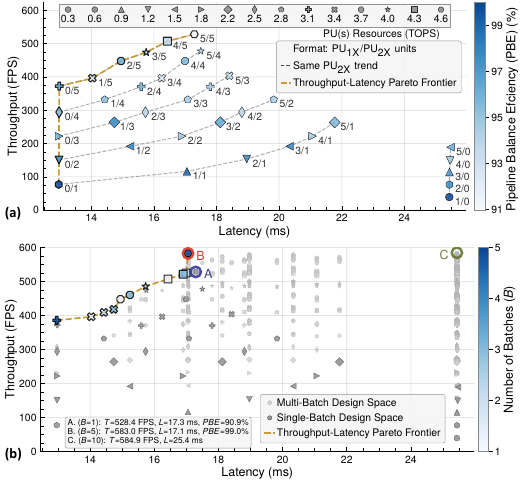}
\vspace{-0.75cm}
\caption{Throughput-latency trade-offs: (a) Single-batch. (b) Multi-batch.}
\label{fig:eval}
\vspace{-0.5cm}
\end{figure}

We selected ResNet-50 \cite{He_2016_CVPR} as our benchmark model, due to its diverse layer composition, which is recognized as a standard benchmark for assessing accelerators' performance by effectively representing various operational characteristics of DNN layers. In Fig.~\ref{fig:eval}(a), the single-batch configurations results are presented, showing the throughput-latency trade-offs, where the throughput is measured in Frames per Second (FPS). The resulting design space demonstrates diversity in performance characteristics, with throughput ranging from configurations utilizing minimal PU resources (marker-coded according to their TOPS utilization), suitable for resource-constrained scenarios, to maximum resources deployment achieving peak performance. The pipeline balance efficiency (PBE, color-coded), calculated consistently with the balance factor equations used in \cite{wu2019compute}, reflects the effectiveness of the node-to-PU assignment algorithm. Single-PU configurations naturally achieve optimal PBE since pipeline coordination is not required. In contrast, multi-PU configurations exhibit varying levels of this metric, depending on the distribution of DNN nodes across PUs and the heterogeneity of the PUs. The throughput-latency Pareto frontier is also shown, establishing the optimal performance boundary and guiding configuration selection based on application requirements.

\begin{table*}[!t]
\centering
\caption{Performance Comparison of FPGA-Based Accelerator Designs on ResNet-50 with INT8 arithmetic.}
\label{tab:fpga_comparison}
\vspace{-0.25cm}
\scriptsize
\renewcommand{\arraystretch}{1.0}
\begin{tabularx}{\textwidth}{@{}l
    >{\raggedright\arraybackslash\tiny}p{1.38cm}   
    S[table-format=4]                              
    S[table-format=3]                              
    r                                              
    S[table-format=2]                              
    S[table-format=2.2]                            
    S[table-format=4.1,group-separator={,},group-minimum-digits=4] 
    S[table-format=4.1,group-separator={,},group-minimum-digits=4] 
    S[table-format=2.1]                            
    S[table-format=1.2]                            
    S[table-format=3.1]                            
    S[table-format=2.1]                            
    S[table-format=2.1]                            
    S[table-format=2.1]                            
    S[table-format=2.3]                            
@{}}
\toprule
\textsc{Architecture} & \scriptsize{FPGA Device} & {Year} &
{\makecell{Freq.\\(\si{\mega\hertz})}} &
{\makecell{Used DSPs\\(Util.\,\%)}} & {\makecell{Batch\\Size}} &
{\makecell{Latency\\(\si{\milli\second})}} & {FPS} & {GOPS} & {CE\,(\%)} &
{GOPS/DSP} & {FPS/TOPS} & {\makecell{Power\\(\si{\watt})}} &
{FPS/W} & {GOPS/W} & {\makecell{Peak\\TOPS}} \\
\midrule
DPU~\cite{amd_pg367,amd_u50_vitis_ai} & XCU50\textsuperscript{1}            & 2021 & 600 & \dspu{3406}{57.2} & 6  & \multicolumn{1}{c}{--} & 572.7 & \grayit{4409.8} & \grayit{59.8} & \grayit{1.29} &  77.7 & \multicolumn{1}{c}{--} & \multicolumn{1}{c}{--} & \multicolumn{1}{c}{--} &  7.373 \\
ShortcutFuse\textsuperscript{3}~\cite{nguyen2022shortcutfusion} & XCKU115\textsuperscript{1} & 2022 & 200 & \dspu{2240}{40.6} &  1 &  11.69 & 85.5 & 1006.0 & 56.1 & 0.45 & 47.7 & \multicolumn{1}{c}{--} & \multicolumn{1}{c}{--} & \multicolumn{1}{c}{--} &  1.792 \\
Full‑Stack~\cite{liu2021toward}      & Arria 10 GX1150\textsuperscript{2} & 2022 & 200 & \dspu{1473}{97.0} &  1 &  5.07 & \grayit{197.3} & 1519.0 & 92.7 & 1.03 & \grayit{120.4} & 19.1 & \grayit{10.3} & 79.5 &  1.638 \\
Rotated~\cite{Fan2022Accel}          & XCKU15P\textsuperscript{1}          & 2022 & 500 & \dspu{1280}{65.0} &  1 &  4.13 & 242.1 & 1874.0 & 73.2 & 1.46 &  94.6 & \multicolumn{1}{c}{--} & \multicolumn{1}{c}{--} & \multicolumn{1}{c}{--} &  2.560 \\
xDNN\cite{d2022xdnn}               & XCU250\textsuperscript{1}          & 2022 & 800 & \dspu{7548}{61.4} &  4 &  3.12 & 1281.0 & \grayit{9863.7} & \grayit{50.2} & \grayit{1.31} & \grayit{65.2} & 128.9 &  9.9 & \grayit{76.5} & \grayit{19.661} \\
Unified Acc.~\cite{li2023unified}  & XCVU37P\textsuperscript{1}          & 2023 & 200 & \dspu{1024}{11.3} &  1 & 13.12 &  76.2 &  590.0 & 72.0 & 0.58 &  93.0 & 17.6 &  4.3 & 33.5 &  0.819 \\
Amoeba~\cite{Wu2024Amoeba}           & Arria 10 SoC\textsuperscript{2}   & 2024 & 200 & \dspu{522}{30.0}  &  1 & 28.03 &  \grayit{35.7} &  286.2 & 69.9 & 0.55 & \grayit{87.2} &  8.2 &  \grayit{4.4} & 35.0 & \grayit{0.410} \\
PipeFuser~\cite{zhou2024pipefuser}   & XCU200\textsuperscript{1}           & 2024 & 220 & \dspu{3560}{52.0} &  1 & \multicolumn{1}{c}{--} & \multicolumn{1}{c}{--} & 2106.0 & \multicolumn{1}{c}{--} & 0.59 & \multicolumn{1}{c}{--} & \multicolumn{1}{c}{--} & \multicolumn{1}{c}{--} & \multicolumn{1}{c}{--} & \multicolumn{1}{c}{--} \\
DCP~\cite{dai2024dcp}               & Stratix 10 GX650\textsuperscript{2} & 2025 & 200 & \dspu{1024}{88.9} &  1 &  9.60 & \grayit{103.9} &  800.0 & 97.7 & 0.78 & \grayit{126.9} &  9.0 & \grayit{11.5} & 88.9 &  0.819 \\
DP‑C                                & XCU50\textsuperscript{1}            & 2025 & 600 & \dspu{3860}{64.8} & 10 & 25.40 & 584.9 & 4515.4 & 98.0 & 1.17 & 126.9 & 46.0 & 12.7 & 98.2 &  4.608 \\
DP‑B                                & XCU50\textsuperscript{1}            & 2025 & 600 & \dspu{3860}{64.8} &  5 & 17.10 & 583.0 & 4500.7 & 97.7 & 1.16 & 126.5 & 46.0 & 12.7 & 97.8 &  4.608 \\
DP‑A                                & XCU50\textsuperscript{1}            & 2025 & 600 & \dspu{3860}{64.8} &  1 & 17.30 & 528.4 & 4079.2 & 88.5 & 1.06 & 114.7 & 46.0 & 11.5 & 88.7 &  4.608 \\
\bottomrule
\end{tabularx}
\vspace{-0.2cm}
\begin{flushleft}
\scriptsize
The \textcolor{GrayCustom}{gray-colored} values are inferred to the best of our understanding from the respective works; Peak TOPS are the DSP TOPS, where for xDNN~\cite{d2022xdnn}, the value uses only the "big block" SAs (6,144 DSPs); FPS/TOPS uses the Peak TOPS. \textsuperscript{1}\,AMD/Xilinx devices. \textsuperscript{2}\,Intel devices. \textsuperscript{3}\,Input Size: 256$\times$256, while the other designs 224$\times$224.
\end{flushleft}
\vspace{-0.65cm}
\end{table*}

Multi-batch configuration analysis, shown in Fig.~\ref{fig:eval}(b), extends the design space to concurrent processing of multiple input streams/batches. In conjunction with the Pareto frontier (applied with a small tolerance), the results reveal how hybrid parallelism strategies can outperform both pure pipeline and pure batch-level approaches, validating the flexibility of our architecture. Also, we highlight three design points that illustrate our architecture's versatility across different deployment scenarios. Design point A (DP-A) represents the highest single-batch throughput achieved by utilizing pipeline parallelism across all PUs, resulting in 90.9\% PBE. In contrast, DP-B achieves maximum system throughput through hybrid parallelism across five concurrent batches, resulting in a 99\% system-level PBE, which demonstrates a key insight. In particular, the flexibility to assign different single-batch configurations to individual pipelines enables better matching of the PU capabilities to DNN layers computational requirements. Despite DP-C using twice as many batches as DP-B, it achieves the same throughput performance, representing maximum batch-level parallelism with one PU per batch, and matching the hardware baseline configuration \cite{petropoulos2025}. Compared to DP-A, it provides a marginal $1.1\times$ throughput improvement with $1.5\times$ latency penalty, demonstrating the significance of DP-B's hybrid parallelism strategy, which balances throughput gains with latency and batch constraints.

Table~\ref{tab:fpga_comparison} presents a comprehensive comparison of our architecture against a broad range of prior FPGA accelerators that optimize DNN inference from different aspects, on ResNet-50 with INT8 arithmetic, encompassing designs for both AMD and Intel platforms. Although different accelerators utilize diverse FPGA devices and hardware resources, we employ normalized metrics to enable the fairest possible comparison. The subsequent analysis focuses on the DP-B configuration, as it exhibits better performance compared to DP-A and DP-C.

DP-B demonstrates competitive performance across multiple metrics, achieving $1.0\times$\,\text{--}\,$2.7\times$ improvement in FPS/TOPS compared to prior works, reflecting each architecture's throughput efficiency in utilizing the DSP resources for the selected target. Energy efficiency analysis reveals that DP-B achieves gains of $1.0\times$\,\text{--}\,$2.9\times$ over prior designs. Notably, the reported power consumption of $46$\,W for both DP-A and DP-B configurations is derived from the hardware baseline average power reported in \cite{petropoulos2025}, assuming comparable power consumption due to the same underlying architectural foundation. The performance density in GOPS/DSP ranges from $0.8\times$ to $2.6\times$ across comparisons. In addition, the compute efficiency (CE), calculated as the ratio of measured GOPS to the available DSP GOPS in each architecture, consistent with the calculations in the compared works, demonstrates improvements of $1.0\times$\,\text{--}\,$1.9\times$, with DP-B reaching 97.7\%. However, system latency presents a trade-off, with DP-B exhibiting $0.6\times$\,\text{--}\,$5.5\times$ performance relative to other designs, i.e., up to $5.5\times$ slower than the fastest implementation, since our architecture is not optimized for low-latency inference. The latter reflects our architectural focus on inter-PU coordination flexibility over intra-layer parallelism optimization, resulting in higher latencies compared to low-latency designs.

\vspace{-0.1cm}
\subsection{Discussion}
\vspace{-0.05cm}
Although our evaluation demonstrated performance improvement over prior works and runtime adaptability, limitation areas can be identified. First, the current approach assigns topologically contiguous DNN layers to PUs, potentially achieving a lower PBE compared to methods that explore non-contiguous subgraph mappings \cite{shen2017maximizing}. Also, the layer‑by‑layer schedule could benefit from tile-triggered pipelines \cite{wei2018tgpa,cai2022deepburning}, where each layer is split into spatial tiles, and the downstream PU starts as soon as the upstream PU has enough activation data. This fine‑grained producer–consumer overlap allows adjacent layers to be launched earlier, thereby reducing latency.

Our framework's multiple phases and the ISA design are orthogonal to these limitations and could enable future integration of such enhancements without fundamental architectural changes. Also, the system architecture can support architectural evolution through PU substitution, where specialized PUs targeting specific layers can replace current PUs while maintaining identical HBM interfaces and coordination mechanisms. In this context, our framework could also accommodate heterogeneous PU configurations, loaded via FPGA reconfiguration, according to specific DNN model requirements.

While ResNet-50 provided detailed insights, the results cannot be generalized to most DNN models, which may exhibit different layer characteristics. The 10-PU design reflects current hardware capabilities rather than inherent coordination constraints. Therefore, our instruction-based coordination can be a viable approach for FPGA acceleration architectures, since programmable inter-PU synchronization enables deployment flexibility while maintaining competitive performance across diverse execution scenarios.

\vspace{-0.1cm}
\section{Conclusion}
\vspace{-0.05cm}
This work introduced an instruction-based coordination architecture that enables runtime reconfiguration between pipeline and hybrid parallelism strategies on the FPGA accelerator, without requiring hardware reprogramming. Our approach combines an ICU per PU with peer-to-peer ISUs, supporting dynamic switching via instruction updates. To enable this coordination, we developed a compilation framework that transforms DNN models into optimized instruction sequences. Experimental validation on ResNet-50 demonstrates CE of $88.5$\%\,\text{--}\,$98.0$\% with higher throughput efficiency ($1.0\times$\,\text{--}\,$2.7\times$ in FPS/TOPS) compared to prior works, while providing deployment flexibility that enables the same architecture to adapt from single-batch pipelining to hybrid parallelism processing.

\section*{Acknowledgment}
This work has been performed in the framework of the EU project "NeuroSoC: A multiprocessor system on chip with in-memory neural processing unit", HORIZON-101070634 \cite{neurosoc}.

\balance
\bibliographystyle{IEEEtran}
\bibliography{IEEEabrv,cites_cooperative}

\begin{thebibliography}{10}
\providecommand{\url}[1]{#1}
\csname url@samestyle\endcsname
\providecommand{\newblock}{\relax}
\providecommand{\bibinfo}[2]{#2}
\providecommand{\BIBentrySTDinterwordspacing}{\spaceskip=0pt\relax}
\providecommand{\BIBentryALTinterwordstretchfactor}{4}
\providecommand{\BIBentryALTinterwordspacing}{\spaceskip=\fontdimen2\font plus
\BIBentryALTinterwordstretchfactor\fontdimen3\font minus \fontdimen4\font\relax}
\providecommand{\BIBforeignlanguage}[2]{{%
\expandafter\ifx\csname l@#1\endcsname\relax
\typeout{** WARNING: IEEEtran.bst: No hyphenation pattern has been}%
\typeout{** loaded for the language `#1'. Using the pattern for}%
\typeout{** the default language instead.}%
\else
\language=\csname l@#1\endcsname
\fi
#2}}
\providecommand{\BIBdecl}{\relax}
\BIBdecl

\bibitem{nechi2023fpga}
A.~Nechi, L.~Groth, S.~Mulhem, F.~Merchant, R.~Buchty, and M.~Berekovic, ``{FPGA-based Deep Learning Inference Accelerators: Where Are We Standing?}'' \emph{{ACM} Trans. Reconfigurable Technol. Syst.}, vol.~16, no.~4, pp. 1--32, Oct. 2023.

\bibitem{zeng2021unified}
S.~Zeng, G.~Dai, H.~Sun, J.~Liu, S.~Li, G.~Ge, K.~Zhong, K.~Guo, Y.~Wang, and H.~Yang, ``{A Unified FPGA Virtualization Framework for General-Purpose Deep Neural Networks in the Cloud},'' \emph{{ACM} Trans. Reconfigurable Technol. Syst.}, vol.~15, no.~3, pp. 1--31, Dec. 2021.

\bibitem{ma2018automatic}
Y.~Ma, Y.~Cao, S.~Vrudhula, and J.-s. Seo, ``{Automatic Compilation of Diverse CNNs Onto High-Performance FPGA Accelerators},'' \emph{{IEEE} Trans. Comput.-Aided Design Integr. Circuits Syst.}, vol.~39, no.~2, pp. 424--437, Feb. 2020.

\bibitem{zhang2018dnnbuilder}
X.~Zhang, J.~Wang, C.~Zhu, Y.~Lin, J.~Xiong, W.-m. Hwu, and D.~Chen, ``{DNNBuilder: an Automated Tool for Building High-Performance DNN Hardware Accelerators for FPGAs},'' in \emph{Proc. {IEEE/ACM} Int. Conf. Comput.-Aided Des. ({ICCAD})}, Nov. 2018, pp. 1--8.

\bibitem{ye2020hybriddnn}
H.~Ye, X.~Zhang, Z.~Huang, G.~Chen, and D.~Chen, ``{HybridDNN: A Framework for High-Performance Hybrid DNN Accelerator Design and Implementation},'' in \emph{Proc. 57th {ACM/IEEE} Des. Automat. Conf. ({DAC})}, Jul. 2020, pp. 1--6.

\bibitem{d2022xdnn}
P.~D'Alberto, V.~Wu, A.~Ng, R.~Nimaiyar, E.~Delaye, and A.~Sirasao, ``{xDNN: Inference for Deep Convolutional Neural Networks},'' \emph{{ACM} Trans. Reconfigurable Technol. Syst.}, vol.~15, no.~2, pp. 1--29, Jan. 2022.

\bibitem{zhou2024pipefuser}
X.~Zhou, S.~Li, H.~Lu, and K.~Wang, ``{PipeFuser: Building Flexible Pipeline Architecture for DNN Accelerators via Layer Fusion},'' in \emph{Proc. 29th Asia and South Pacific Des. Automat. Conf. ({ASP-DAC})}, Jan. 2024, pp. 921--926.

\bibitem{zhang2020dnnexplorer}
X.~Zhang, H.~Ye, J.~Wang, Y.~Lin, J.~Xiong, W.-m. Hwu, and D.~Chen, ``{DNNExplorer: A Framework for Modeling and Exploring a Novel Paradigm of FPGA-based DNN Accelerator},'' in \emph{Proc. {IEEE/ACM} Int. Conf. Comput.-Aided Des. ({ICCAD})}, Dec. 2020, pp. 1--9.

\bibitem{basalama2023flexcnn}
S.~Basalama, A.~Sohrabizadeh, J.~Wang, L.~Guo, and J.~Cong, ``{FlexCNN: An End-to-end Framework for Composing CNN Accelerators on FPGA},'' \emph{{ACM} Trans. Reconfigurable Technol. Syst.}, vol.~16, no.~2, pp. 1--32, Mar. 2023.

\bibitem{xing2019dnnvm}
Y.~Xing \emph{et~al.}, ``{DNNVM: End-to-End Compiler Leveraging Heterogeneous Optimizations on FPGA-Based CNN Accelerators},'' \emph{{IEEE} Trans. Comput.-Aided Design Integr. Circuits Syst.}, vol.~39, no.~10, pp. 2668--2681, Oct. 2020.

\bibitem{li2025hardware}
J.~Li, W.~Wang, and W.-J. Li, ``{Hardware Computation Graph for DNN Accelerator Design Automation Without Inter-PU Templates},'' \emph{{IEEE} Trans. Comput.-Aided Design Integr. Circuits Syst.}, vol.~44, no.~11, pp. 4276--4289, Nov. 2025.

\bibitem{cai2022deepburning}
X.~Cai, Y.~Wang, X.~Ma, Y.~Han, and L.~Zhang, ``{DeepBurning-SEG: Generating DNN Accelerators of Segment-Grained Pipeline Architecture},'' in \emph{Proc. 55th {IEEE/ACM} Int. Symp. Microarchit. ({MICRO})}, Oct. 2022, pp. 1396--1413.

\bibitem{amd_dsp48e2_guide}
\BIBentryALTinterwordspacing
{AMD, Inc., Santa Clara, CA, USA}, ``{UltraScale Architecture DSP Slice (UG579)},'' 2021. [Online]. Available: \url{https://docs.amd.com/v/u/en-US/ug579-ultrascale-dsp}
\BIBentrySTDinterwordspacing

\bibitem{shi2020ftdl}
R.~Shi \emph{et~al.}, ``{FTDL: A Tailored FPGA-Overlay for Deep Learning with High Scalability},'' in \emph{Proc. 57th {ACM/IEEE} Des. Automat. Conf. ({DAC})}, Jul. 2020, pp. 1--6.

\bibitem{Fan2022Accel}
X.~Fan, G.~Xie, Z.~Huang, W.~Cao, and L.~Wang, ``{Acceleration of Rotated Object Detection on FPGA},'' \emph{{IEEE} Trans. Circuits Syst. {II}}, vol.~69, no.~4, pp. 2296--2300, Apr. 2022.

\bibitem{petropoulos2025}
A.~Petropoulos and T.~Antonakopoulos, ``{A Scalable FPGA Architecture With Adaptive Memory Utilization for GEMM-Based Operations},'' \emph{{IEEE} Trans. {VLSI} Syst.}, vol.~33, no.~8, pp. 2334--2338, Aug. 2025.

\bibitem{li2023unified}
T.~Li, F.~Zhang, X.~Fan, J.~Shen, W.~Guo, and W.~Cao, ``{Unified Accelerator for Attention and Convolution in Inference Based on FPGA},'' in \emph{Proc. {IEEE} Int. Symp. Circuits Syst. ({ISCAS})}, May 2023, pp. 1--5.

\bibitem{zhang2024visionagile}
B.~Zhang, R.~Kannan, C.~Busart, and V.~K. Prasanna, ``{VisionAGILE: A Versatile Domain-Specific Accelerator for Computer Vision Tasks},'' \emph{{IEEE} Trans. Parallel Distrib. Syst.}, vol.~35, no.~12, pp. 2405--2422, Dec. 2024.

\bibitem{Wu2024Amoeba}
X.~Wu, M.~Wang, J.~Lin, and Z.~Wang, ``{Amoeba: An Efficient and Flexible FPGA-Based Accelerator for Arbitrary-Kernel CNNs},'' \emph{{IEEE} Trans. {VLSI} Syst.}, vol.~32, no.~6, pp. 1086--1099, Jun. 2024.

\bibitem{dai2024dcp}
K.~Dai, Z.~Xie, and S.~Liu, ``{DCP-CNN: Efficient Acceleration of CNNs With Dynamic Computing Parallelism on FPGA},'' \emph{{IEEE} Trans. Comput.-Aided Design Integr. Circuits Syst.}, vol.~44, no.~2, pp. 540--553, Feb. 2025.

\bibitem{amd_pg367}
\BIBentryALTinterwordspacing
{AMD, Inc., Santa Clara, CA, USA}, ``{DPUCAHX8H Performance (PG367)},'' 2024. [Online]. Available: \url{https://docs.amd.com/r/en-US/pg367-dpucahx8h/Performance}
\BIBentrySTDinterwordspacing

\bibitem{xiao2017exploring}
Q.~Xiao, Y.~Liang, L.~Lu, S.~Yan, and Y.-W. Tai, ``{Exploring heterogeneous algorithms for accelerating deep convolutional neural networks on FPGAs},'' in \emph{Proc. 54th {ACM/IEEE} Des. Automat. Conf. ({DAC})}, Jun. 2017, pp. 1--6.

\bibitem{shen2017maximizing}
Y.~Shen, M.~Ferdman, and P.~Milder, ``{Maximizing CNN Accelerator Efficiency Through Resource Partitioning},'' in \emph{Proc. {ACM/IEEE} 44th Annu. Int. Symp. Comput. Archit. ({ISCA})}, Jun. 2017, pp. 535--547.

\bibitem{wu2019compute}
E.~Wu, X.~Zhang, D.~Berman, I.~Cho, and J.~Thendean, ``{Compute-Efficient Neural-Network Acceleration},'' in \emph{Proc. {ACM/SIGDA} Int. Symp. Field-Program. Gate Arrays ({FPGA})}, Feb. 2019, pp. 191--200.

\bibitem{chan2019partitioning}
L.~C. Chan, G.~Malik, and N.~Kapre, ``{Partitioning FPGA-Optimized Systolic Arrays for Fun and Profit},'' in \emph{Proc. Int. Conf. Field-Program. Technol. ({ICFPT})}, Dec. 2019, pp. 144--152.

\bibitem{doumet2024h2pipe}
M.~Doumet, M.~Stan, M.~Hall, and V.~Betz, ``{H2PIPE: High Throughput CNN Inference on FPGAs with High-Bandwidth Memory},'' in \emph{Proc. 34th Int. Conf. Field Program. Log. Appl. ({FPL})}, Sep. 2024, pp. 69--77.

\bibitem{wei2018tgpa}
X.~Wei, Y.~Liang, X.~Li, C.~H. Yu, P.~Zhang, and J.~Cong, ``{TGPA: Tile-Grained Pipeline Architecture for Low Latency CNN Inference},'' in \emph{Proc. {IEEE/ACM} Int. Conf. Comput.-Aided Des. ({ICCAD})}, Nov. 2018, pp. 1--8.

\bibitem{samajdar2019scaling}
A.~Samajdar, T.~Garg, T.~Krishna, and N.~Kapre, ``{Scaling the Cascades: Interconnect-Aware FPGA Implementation of Machine Learning Problems},'' in \emph{Proc. 29th Int. Conf. Field-Program. Log. Appl. ({FPL})}, Sep. 2019, pp. 342--349.

\bibitem{abdelfattah2018dla}
M.~S. Abdelfattah \emph{et~al.}, ``{DLA: Compiler and FPGA Overlay for Neural Network Inference Acceleration},'' in \emph{Proc. 28th Int. Conf. Field-Program. Log. Appl. ({FPL})}, Aug. 2018, pp. 411--418.

\bibitem{yu2019opu}
Y.~Yu, C.~Wu, T.~Zhao, K.~Wang, and L.~He, ``{OPU: An FPGA-Based Overlay Processor for Convolutional Neural Networks},'' \emph{{IEEE} Trans. {VLSI} Syst.}, vol.~28, no.~1, pp. 35--47, Jan. 2020.

\bibitem{onnx}
\BIBentryALTinterwordspacing
{ONNX Community}, ``{Open Neural Network Exchange (ONNX)},'' 2024. [Online]. Available: \url{https://github.com/onnx/onnx/releases/tag/v1.16.0}
\BIBentrySTDinterwordspacing

\bibitem{toupas2024}
P.~Toupas, Z.~Yu, C.-S. Bouganis, and D.~Tzovaras, ``{SMOF: Streaming Modern CNNs on FPGAs with Smart Off-Chip Eviction},'' in \emph{Proc. {IEEE} 32nd Annu. Int. Symp. Field-Program. Custom Comput. Mach. ({FCCM})}, May 2024, pp. 185--196.

\bibitem{huang2021shuhai}
H.~Huang \emph{et~al.}, ``{Shuhai: A Tool for Benchmarking High Bandwidth Memory on FPGAs},'' \emph{{IEEE} Trans. Comput.}, vol.~71, no.~5, pp. 1133--1144, May 2022.

\bibitem{He_2016_CVPR}
K.~He, X.~Zhang, S.~Ren, and J.~Sun, ``{Deep Residual Learning for Image Recognition},'' in \emph{Proc. {IEEE} Conf. Comput. Vis. Pattern Recognit. ({CVPR})}, Jun. 2016, pp. 770--778.

\bibitem{amd_u50_vitis_ai}
\BIBentryALTinterwordspacing
{AMD, Inc., Santa Clara, CA, USA}, ``{Vitis AI (UG1354)},'' 2021. [Online]. Available: \url{https://docs.amd.com/r/1.4.1-English/ug1354-xilinx-ai-sdk/Alveo-U50/U50LV-Data-Accelerator-Card}
\BIBentrySTDinterwordspacing

\bibitem{nguyen2022shortcutfusion}
D.~T. Nguyen, H.~Je, T.~N. Nguyen, S.~Ryu, K.~Lee, and H.-J. Lee, ``{ShortcutFusion: From Tensorflow to FPGA-Based Accelerator With a Reuse-Aware Memory Allocation for Shortcut Data},'' \emph{{IEEE} Trans. Circuits Syst. {I}}, vol.~69, no.~6, pp. 2477--2489, Jun. 2022.

\bibitem{liu2021toward}
S.~Liu, H.~Fan, M.~Ferianc, X.~Niu, H.~Shi, and W.~Luk, ``{Toward Full-Stack Acceleration of Deep Convolutional Neural Networks on FPGAs},'' \emph{{IEEE} Trans. Neural Netw. Learn. Syst.}, vol.~33, no.~8, pp. 3974--3987, Aug. 2022.

\bibitem{neurosoc}
\BIBentryALTinterwordspacing
NeuroSoC. [Online]. Available: \url{https://neurosoc.eu/}
\BIBentrySTDinterwordspacing

\end{thebibliography}

\end{document}